# AcerDET[1] : a particle level fast simulation and reconstruction package for phenomenological studies on high $p_T$ physics at LHC.

**Elżbieta Richter-Wąs**[2]
*Institute of Computer Science, Jagellonian University;*
*30-072 Krakow, ul. Nawojki 11, Poland.*
*Institute of Nuclear Physics*
*31-342 Krakow, ul. Radzikowskiego 152, Poland.*

**Abstract**

This paper documents package for the particle level fast simulation. The package is designed to complete the AcerMC generator framework with the easy-to-use simulation and reconstruction algorithms. The package provides, starting from list of particles in the event, the list of reconstructed jets, isolated electrons, muons and photons and reconstructed missing transverse energy. The AcerDET represents a simplified version of the package called ATLFAST, used since several years within ATLAS Collaboration. In the AcerDET version some functionalities of the former one have been removed, only the most crucial detector effects are implemented and the parametrisations are largely simplified. Therefore it is not representing in details neither ATLAS nor CMS detectors. Nevertheless, we believe that the package can be well adequate for some feasibility studies of the high $p_T$ physics at LHC and in future, after some adjustments, of other detectors as well.

[1]Package is available for the web page `http://erichter.home.cern.ch/erichter/AcerDET.html`
[2] Supported in part by Polish Government grant KBN 2 P03B 001 22.

PROGRAM SUMMARY

*Title of the program:* **AcerDET version 1.0**

*Operating system:* Linux

*Programming language:* FORTRAN 77 with popular extensions.

*External libraries:* CERNLIB.

*Size of the compressed distribution directory:* about 25 kB.

*Key words:* Fast simulation, Physics at LHC.

*Nature of physical problem:* A particle level fast simulation and reconstruction package. The package provides, starting from the list of particles in the event, list of reconstructed jets, isolated electrons, muons and photons and reconstructed missing transverse energy. The package is aimed to complete the `AcerMC` framework [1] with the easy-to-use simulation and reconstruction algorithms. The interface subroutines to `PYTHIA 6.2` [2] or `HERWIG 6.3` [3] are provided. Distribution version includes also example of the main program for execution with `PYTHIA 6.2` generator. Implemented set of parametrisations is not representing in details ATLAS or CMS detectors, some of them are simple and can be considered rather as place-holders for future adaptation of any detector. Nevertheless, we believe that the package will be well adequate for some feasibility studies on the high $p_T$ physics at LHC.

# 1 Introduction

The potential of the LHC detectors for physics at high $p_T$ will be very rich, see [1] for comprehensive review. Prospects for observability of eg. the Higgs boson(s), Supersymmetry particles, Exotic particles, New Vector Bosons is very impressive. All these thanks to the high sensitivity of the detectors in terms of acceptance and identifying efficiencies to variety of signatures: photons, electrons, muons, multi-b-jets, tau-jets, missing transverse energy.

Those high sensitivities will allow to study very exclusive signatures and so the discovery potential will be limited in some cases by the rare background processes. It is becoming evident that the multi-jet and multi-b-jet production in association with known vector bosons, W and/or Z, or with top-quark pair will be the serious background to the several observables as well. As so, should be understood beyond the leading order. One will likely need to understand well effects coming from the finite width or angular spin correlations (eg. from the intermediate resonances decays). Moreover one should be able to understand limitations and complementarity of the matrix element and parton shower predictions for the diversity of signatures.

The package for particle-level simulation and reconstruction is one of the intermediate steps between simple parton-level analysis and very sophisticated and CPU consuming full detector simulation. The package provides, starting from list of particles in the event, list of reconstructed jets, isolated electrons, muons and photons and reconstructed missing transverse energy. It can serve for several phenomenological feasibility studies on the prospects for observability of a given signature. One of the example of such application reported in [4] are studies on prospects for observability of the invisibly decaying Higgs boson in the $t\bar{t}H$ production process. Our package can be also useful for the dedicated comparisons between matrix element and parton shower predictions. In that case one can compare experimental signatures (reconstructed jets, leptons, photons) to quantify size of the discrepancies between different predictions in straightforward way. Such comparison was recently reported in [5] for $Wb\bar{b}$, $Zb\bar{b}$ and $t\bar{t}$ processes.

The package simulates some key features of the LHC detectors like ATLAS and CMS. It is based on the calorimetric energy deposition for jets reconstruction and tracks reconstruction for electrons and muons. It takes into account very high granularity of the electromagnetic calorimetry for the photon reconstruction. The missing transverse energy is calculated for the total energy balance of the reconstructed objects. The capability for the identification of b-jets and tau-jets is also explored. Implemented set of parametrisations is not representing in details performance of neither ATLAS nor CMS detectors. Nevertheless, we believe that the package will be adequate for some feasibility studies on the high $p_T$ physics at LHC and offer option also for LC.

The paper is organised as follows. In Section 2 we discuss algorithms used for objects reconstruction and show benchmarking distributions. Section 3 gives an outlook. Some comparison numbers with the performance of the ATLAS detector are collected in Appendix A. In the Appendixes B-F we give more technical details concerning input parameters and output structure. In Appendix G we show control output.

## 2 Simulation and reconstruction

Fully or partially generated event i.e. event generated including or not QED/QCD initial and final state radiation, fragmentation, hadronisation and decays of unstable particles can by analysed by this package. The list of the partons/particles of the generated event should be rewritten (including information on their history) from the event record, filled by the event generator, to the `COMMON /ACMCEVENT/`. Stored there information is used by the `AcerDET` algorithms for event simulation and reconstruction.
Events simulation is limited to the following steps:

- Deposition of the particles energies in the calorimetric cells.
- Smearing of the energy of electrons, photons, muons with the parametrised resolutions.
- Smearing of the energy of hadronic clusters and not-clustered cells with the parametrised resolution.

Events reconstruction is limited to the following steps:

- Reconstruction of the calorimetric clusters.
- Verification of the isolation criteria for electrons/photons/muons.
- Rejection of clusters associated with electrons and photons.
- Acceptation of remaining clusters as hadronic jets and identification (labeling) of those associated with b-quarks, c-quarks, tau-leptons.
- Jets energy calibration.

There is no clear separation between the *simulation* and *reconstruction* parts in the structure of algorithms. Some larger blocks (`subroutines`) might realise both tasks.

As a final result provided are four-momenta of reconstructed electrons, photons, muons, labeled and calibrated jets and calculated is total missing transverse energy.

### 2.1 Calorimetric clusters

The transverse energy of all undecayed particles stored in `COMMON /ACMCEVENT/`, except for muons, neutrinos and other *invisible* particles[3] eg. the lightest SUSY particle, are summed up in the map of calorimetric cells with a given granularity in $(\eta \times \phi)$ coordinates (default: $0.1 \times 0.1$ for $|\eta| < 3.2$ and $0.2 \times 0.2$ for $|\eta| > 3.2$, with the calorimetric coverage up to $|\eta| = 5.0$). As an effect of the solenoidal magnetic field in the inner part of the detector we assume that the $\phi$ position of charged particles with transverse momenta above the threshold (default: $p_T > 0.5$ GeV) will be shifted as parametrised in function `FLDPHI`. The contribution from all charged particles with transverse momenta below that threshold is neglected.

[3] User may want to make any particle invisible to the detector. For this one should redefine its code in the `COMMON /ACMCEVENT/` to that specified for the invisible particles in `acerdet.dat` file.

All calorimetric cells with the transverse energy greater than a given threshold (default: $E_T > 1.5$ GeV) are taken as possible initiators of clusters. They are scanned in order of decreasing $E_T$ to verify whether the total $E_T$ summed over all cells in a cone $\Delta R = \sqrt{\Delta^2\eta + \Delta^2\phi}$ exceeds the minimum required threshold for the reconstructed cluster (default $E_T > 5$ GeV). Cells with deposited transverse energy below the threshold (default: $E_T = 0$) are not accounted for. As a coordinates ($\eta^{clu} \times \phi^{clu}$) of the reconstructed cluster taken are the coordinates of the bary-center of the cone weighted by the cells $E_T$ for all cells inside the cone around the initiator cell.

All reconstructed clusters are stored in the `COMMON /CLUSTE/`. Fig. 1 shows the $\Delta\eta$ (left) and $\Delta\phi$ (right) distribution between the reconstructed bary-center of particles falling within the geometrical cluster cone and the reconstructed cluster position.

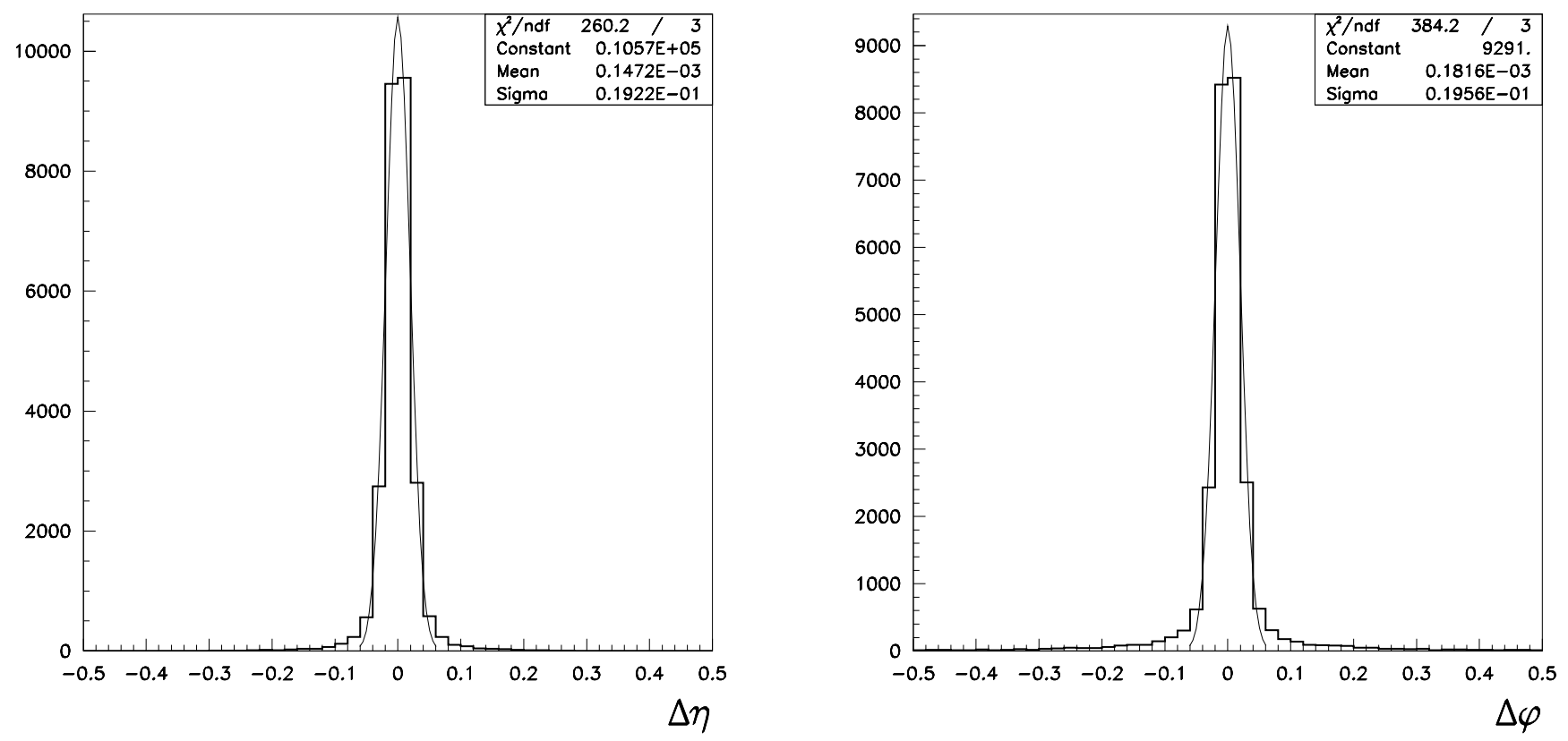

Figure 1: *The $\Delta\eta$ (left) and $\Delta\phi$ (right) distribution between the reconstructed bary-center of particles falling within the cluster cone and the reconstructed cluster position. Shown are results for generated $WH, H \to u\bar{u}$ process with $m_H = 100$ GeV.*

## 2.2 Isolated muons

Algorithm reconstructing isolated muons uses information of the generated muons, reconstructed calorimetric clusters and the cells map.

Isolated muon candidates are searched for in the `COMMON /ACMCEVENT/`. The inverse muon four-momentum is smeared according to the Gaussian resolution parametrised with function `RESMUO` (default: $\sigma = 0.05\% \cdot p_T$). The muon direction remains unsmeared.

For all muons which pass selection criteria in $p_T$ and $\eta$ (default: $p_T > 6$ GeV and $|\eta| < 2.5$ GeV), isolation criteria, in terms of the distance from calorimetric clusters and of maximum transverse energy deposition in cells in a cone around the muon, are then applied (defaults: separation by $\Delta R > 0.4$ from other clusters and $\sum E_T < 10$ GeV in a cone $\Delta R = 0.2$ around the muon). All muons passing the isolation criteria are stored in `COMMON /ISOMUO/` and those not passing are stored in `COMMON /NIOMUO/`.

As a control physics process the $gg \to H \to ZZ^* \to 4\mu$ production with the Higgs boson mass of 130 GeV is used. The default isolation selection has 97.8% efficiency for muons passing kinematical selection with $p_T^{\mu_1,\mu_2} > 20$ GeV and $p_T^{\mu_3,\mu_4} > 7$ GeV. As the predicted intrinsic width of the Higgs boson of this mass is very small, the resolution is completely dominated by the resolution assumed for the single muon reconstruction. Fig. 2 (left) shows the reconstructed distribution of the 4-muon system. The assumed single muon transverse momenta resolution of $\sigma = 0.05\% \cdot p_T$ leads to the $\sigma_m = 1.57$ GeV resolution for the invariant mass of the four-muon system originating from the $H \to ZZ^* \to 4\mu$ decay. Please note that the photon bremsstrahlung was omitted in the event generation.

The second control physics process is the di-jet production. Here we don't expect isolated muons to be present, just muons from the semileptonic cascade within the jets. Nevertheless some of those muons will also pass isolation criteria. Fig. 2 (right) shows the $p_T$ distribution of the true (dashed) and classified as isolated (solid) muons passing the default selection criteria.

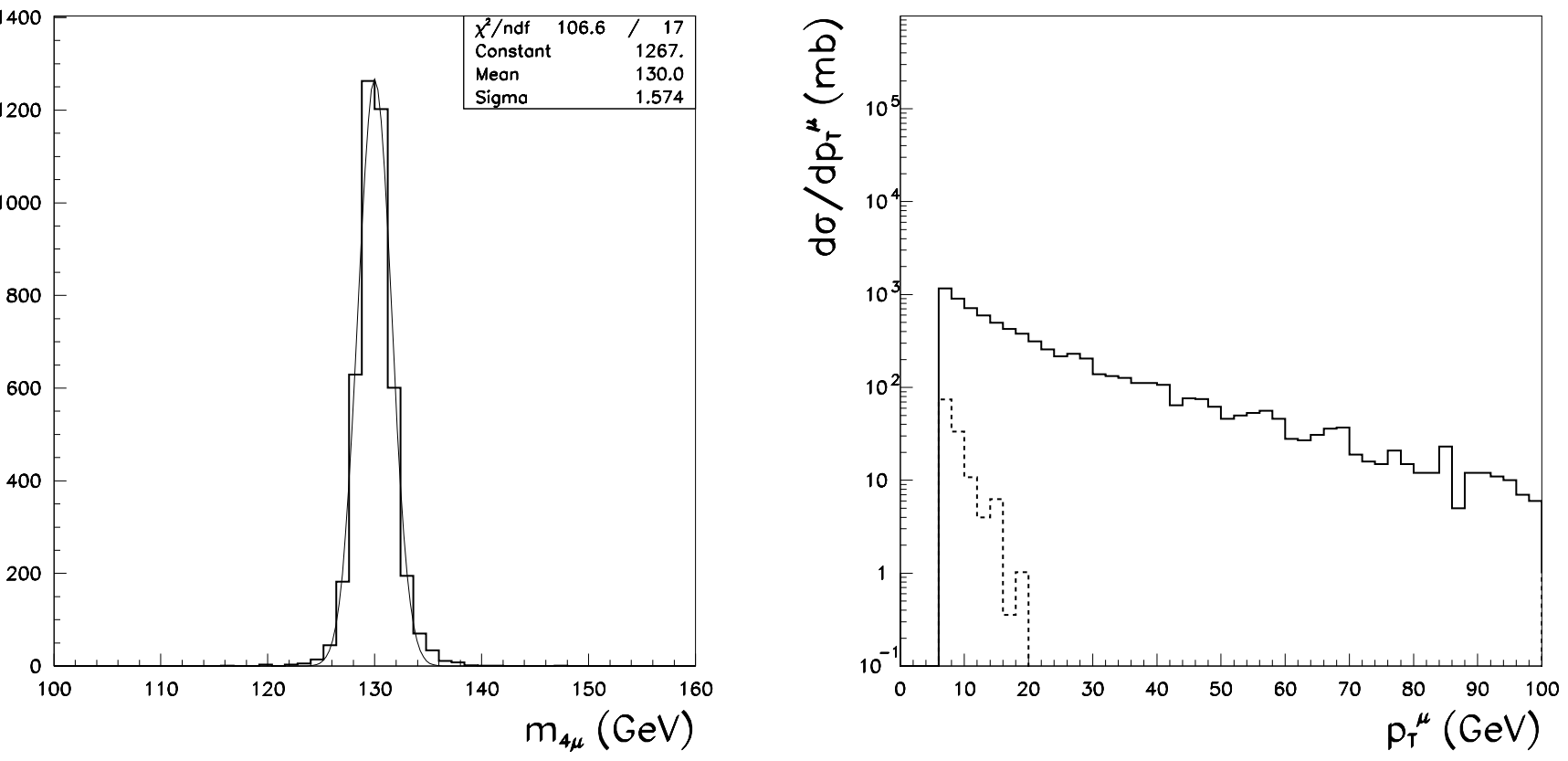

Figure 2: *Left: The reconstructed mass, $m_{4\mu}$, for $H \to ZZ^* \to 4\mu$ events with $m_H = 130$ GeV; Right: the transverse momenta spectrum of true (solid) and reconstructed isolated (dashed) muons in the di-jet sample generated with $p_T^{hard} > 17$ GeV.*

## 2.3 Isolated electrons

Algorithm reconstructing isolated electrons uses information of the generated electrons, reconstructed calorimetric clusters and the cells map.

Isolated electron candidates are searched for in the `COMMON /ACMCEVENT/`. The electron four-momentum is smeared according to the Gaussian resolution parametrised with function `RESELE` (default: $\sigma = 12\%/\sqrt{E}$). The electron direction remains unsmeared.

For all electrons which pass selection criteria in $p_T$ and $\eta$ (default: $p_T > 5$ GeV and $|\eta| < 2.5$ GeV), the associated reconstructed calorimeter cluster is identified (default: $\Delta R_{e,cluster} < 0.1$). Electron isolation criteria, in terms of the distance from other clusters and of maximum transverse energy deposition in cells in a cone around the electron, are then applied (defaults: separation by $\Delta R > 0.4$ from other clusters and $\sum E_T < 10$ GeV in a cone $\Delta R = 0.2$ around the electron). All electrons passing the isolation criteria are stored in COMMON /ISOELE/ and the clusters associated with them are removed from the COMMON /CLUSTE/.

As a control physics process the $gg \to H \to ZZ^* \to 4e$ production with the Higgs boson mass of 130 GeV is used. The default isolation selection has 95.2% efficiency for electrons passing kinematical selection with $p_T^{e_1,e_2} > 20$ GeV and $p_T^{e_3,e_4} > 7$ GeV. As the predicted intrinsic width of the Higgs boson of this mass is very small, the resolution is completely dominated by the resolution assumed for the single electron reconstruction. Fig. 3 (left) shows the reconstructed distribution of the 4-electron system. The assumed single electron energy resolution of $\sigma = 12\%/\sqrt{E}$ leads to the $\sigma_m = 1.26$ GeV resolution for the reconstructed invariant mass of the four-electron system originating from the $H \to ZZ^* \to 4e$ decay. Please note that the photon bremsstrahlung was omitted in the events generation.

The second control physics process is the di-jet production. Here we don't expect isolated electrons to be present, just electrons from the semileptonic cascade or Dalitz decays within the jets. Nevertheless some of these electrons will also pass isolation criteria. Fig. 3 (right) shows the $p_T$ distribution of the true (dashed) and classified as isolated (solid) electrons passing the default selection criteria.

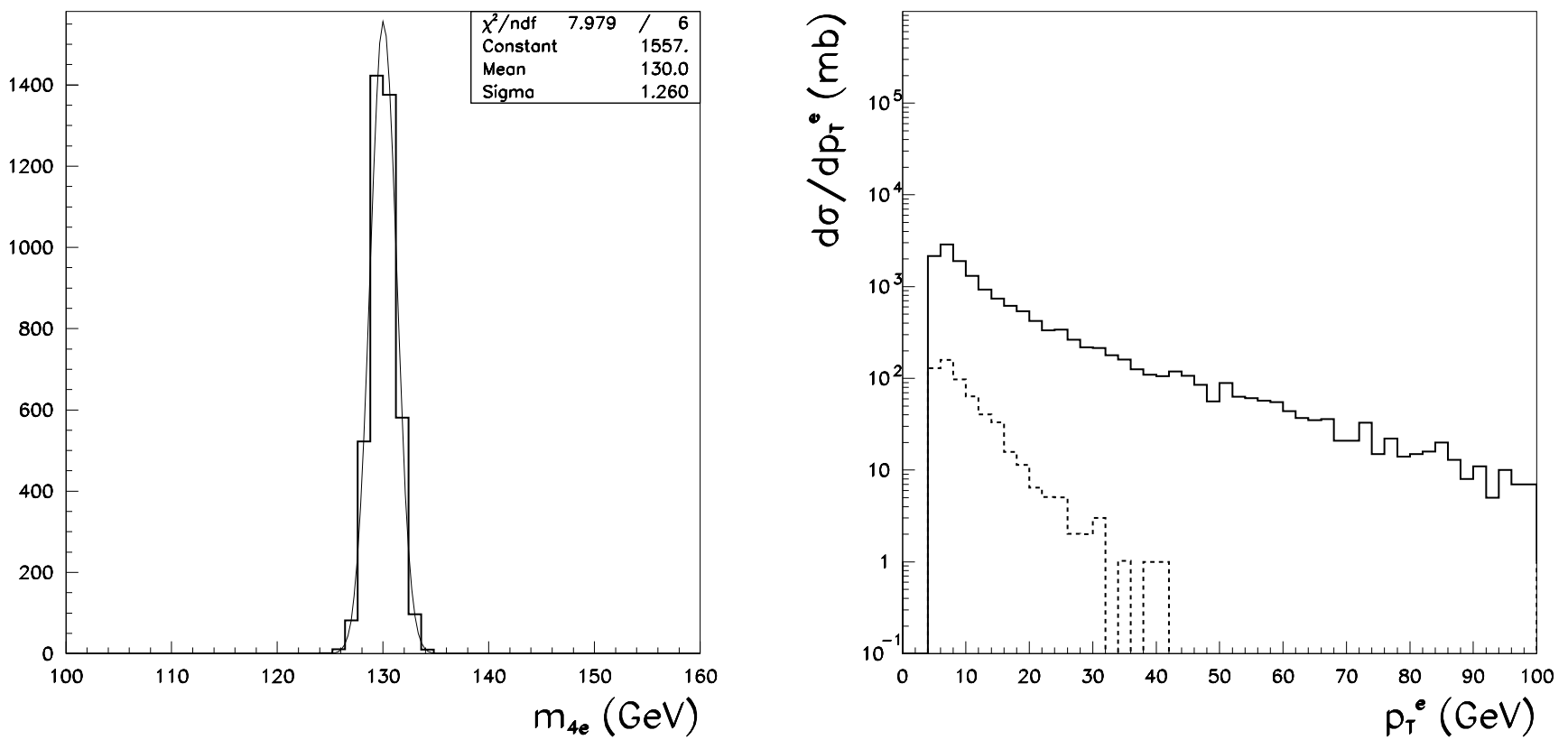

Figure 3: *Left: The reconstructed mass, $m_{4e}$, for $H \to ZZ^* \to 4e$ events with $m_H = 130$ GeV; Right: the transverse momenta spectrum of true (solid) and reconstructed isolated (dashed) electrons in the di-jet sample generated with $p_T^{hard} > 17$ GeV.*

## 2.4 Isolated photons

Algorithm reconstructing isolated photons uses information of the generated photons, reconstructed calorimetric clusters and the cells map.

Isolated photon candidates are searched for in the COMMON /ACMCEVENT/. The photon four-momentum is smeared according to the Gaussian resolution parametrised with function RESPHO (default: $\sigma = 10\%/\sqrt{E}$). The photon direction remains unsmeared.

For all photons which pass selection criteria in $p_T$ and $\eta$ (default: $p_T > 5$ GeV and $|\eta| < 2.5$ GeV), the associated reconstructed calorimeter cluster is identified (default:

$\Delta R_{\gamma,cluster} < 0.1$). Photon isolation criteria, in terms of the distance from other clusters and of maximum transverse energy deposition in cells in a cone around the photon, are then applied (defaults: separation by $\Delta R > 0.4$ from other clusters and $\sum E_T < 10$ GeV in a cone $\Delta R = 0.2$ around the photon). All photons passing the isolation criteria are stored in `COMMON /ISOPHO/` and the clusters associated with them are removed from the `COMMON /CLUSTE/`.

As a control physics process the Standard Model $gg \to H \to \gamma\gamma$ production with the Higgs boson mass of 100 GeV is used. The isolation selection has 98.0% efficiency for photons passing the kinematical selection of $p_T^{\gamma_1} > 40$ GeV and $p_T^{\gamma_2} > 25$ GeV. As the predicted intrinsic width of the Higgs boson of this mass is much below 1 GeV, the resolution of the reconstructed invariant mass of the di-photon system is completely dominated by the resolution assumed for the single photon reconstruction. Fig. 4 (left) shows the reconstructed distribution of the di-photon system of photons passing selection criteria. The assumed energy resolution of $\sigma = 10\%/\sqrt{E}$ leads to the $\sigma_m = 0.85$ GeV resolution for the reconstructed invariant mass of the di-photon system originating from the $H \to \gamma\gamma$ decay.

The second control physics process is the di-jet production. Here we don't expect isolated photons to be present, just photons from the radiative cascade or $\pi's$ decays within the jets. Nevertheless, some of those photons will also pass isolation criteria. Fig. 4 (right) shows the $p_T$ distribution of the true (dashed) and classified as isolated (solid) photons passing the default selection criteria.

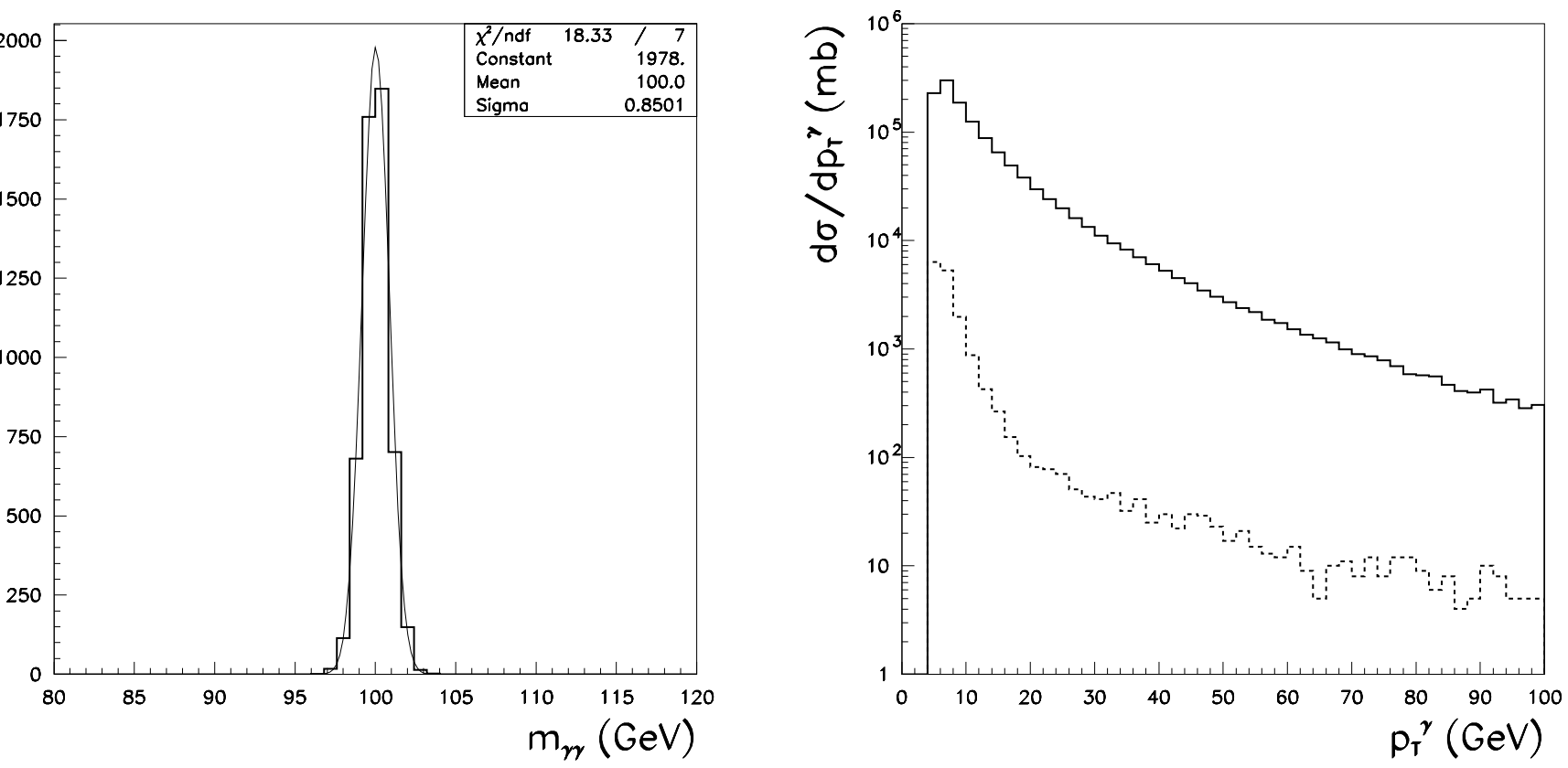

Figure 4: *Left: The reconstructed mass, $m_{\gamma\gamma}$, for $H \to \gamma\gamma$ events with $m_H = 100$ GeV; Right: the transverse momenta spectrum for true (solid) and reconstructed isolated (dashed) photons in the di-jet sample generated with $p_T^{hard} > 17$ GeV.*

## 2.5 Jets

Clusters which have not been selected as associated with electrons or photons are smeared with Gaussian resolution parametrised in function `RESHAD` (default: $\sigma = 50\%/\sqrt{E}$ and $100\%/\sqrt{E}$).

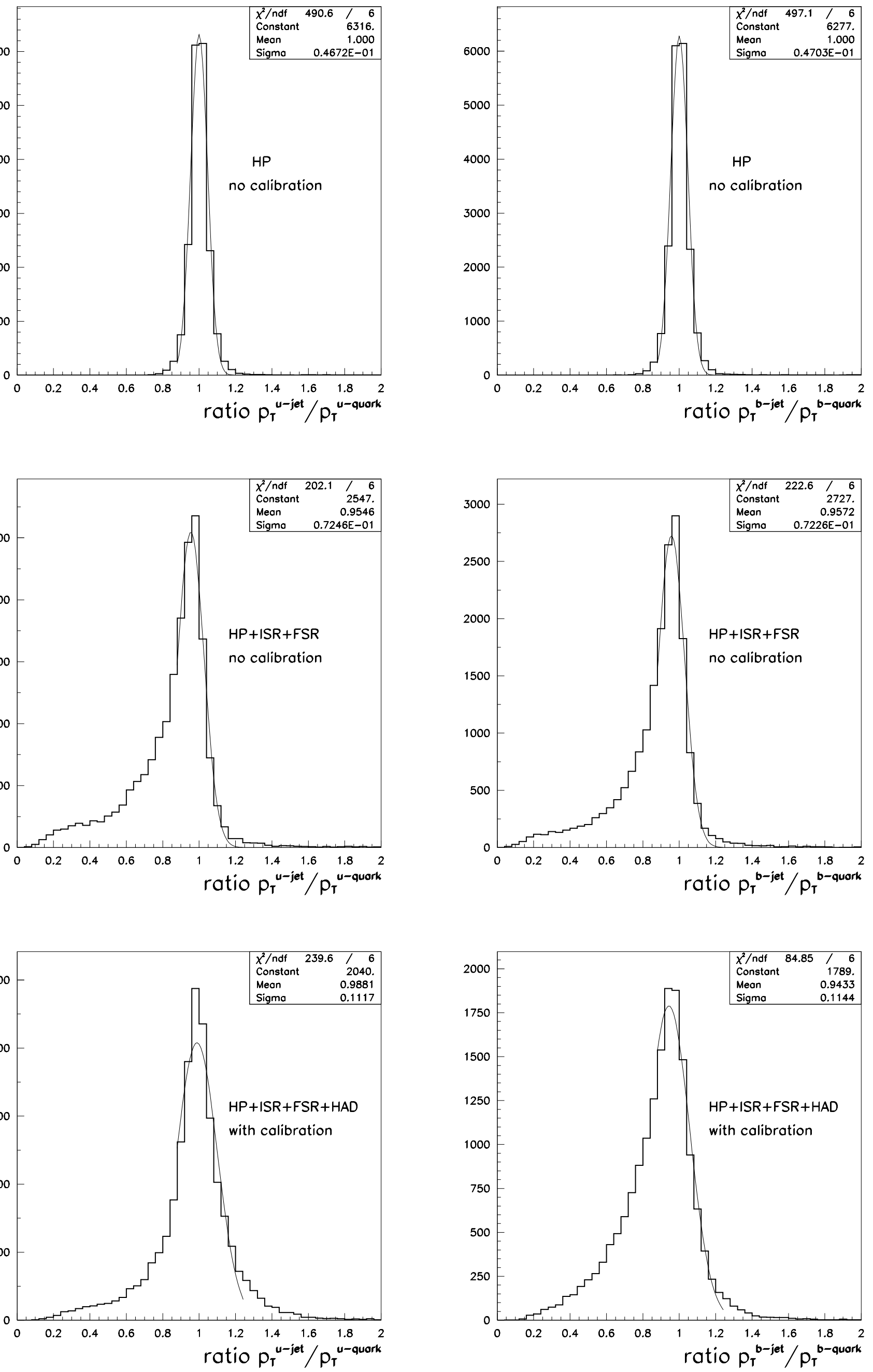

Figure 5: *The ratio of $p_T^{u-jet}/p_T^{u-quark}$ (left) and $p_T^{b-jet}/p_T^{b-quark}$ (right) for events generated with hard process only (top), hard process and ISR/FSR (middle) and full generation (hard process + ISR/FSR + hadronisation) (bottom). Events are calibrated only in case of the full generation (bottom plots). The HP denotes hard process, ISR/FSR initial and final state radiation respectively and HD denotes hadronisation.*

If the non-isolated muon falls into the cone of a cluster its 4-momenta is added to the cluster 4-momenta and the cluster direction is recalculated. The resulting clusters are classified as jets if their transverse momentum is greater than a given threshold (default: $p_T > 15$ GeV). They are removed from the `COMMON /CLUSTE/` and stored in the common `COMMON /JETALL/`.

The jets reconstruction efficiencies and di-jet mass resolution have been studied using control physics process of the $WH$ production with $m_H = 100$ GeV and forcing the Higgs boson decay into specific partons, namely $H \to u\bar{u}$, $H \to c\bar{c}$ and $H \to b\bar{b}$. We used also process $gg \to H \to \tau\tau$ for estimating tau-jet reconstruction efficiency (we forced tau-leptons decay into hadrons). For estimating reconstruction efficiency we consider only jets which have been reconstructed within the cone $\Delta R = 0.4$ from the primary parton (particle) and we require that the primary parton (particle) has passed the same kinematical selection as required for reconstructed jets.

### 2.5.1 Labeling

Very important for the physics at LHC are jets originating from b-quarks (so called b-jets) which can be identified in the detector using b-tagging technique (vertex or soft-lepton tags). The package labels a jet as a b-jet if it is reconstructed within a limited rapidity range (default: $|\Delta\eta| < 2.5$) and if a b-quark of a transverse momenta (after FSR) above the threshold (default: $p_T > 5$ GeV) is found within the cone (default: $\Delta R = 0.2$) around the axis of reconstructed jet. The similar criteria are used for labeling the c-jets.

Equivalently important are also jets originating from the hadronic $\tau$-decay (so called $\tau$-jets) which can be identified using dedicated algorithms. The package labels jet as a tau-jet if the hadronic decay product is relatively hard (default: $p_T^{\tau-had} > 10$ GeV), inside limited rapidity range (default: $|\eta| < 2.5$), dominates reconstructed jet transverse momenta (default: $p_T^{\tau-had}/p_T^{jet} > 0.9$), and is within the cone (default: $\Delta R_{jet,\tau-had} < 0.3$) around the axis of a jet.

Table 1 summarises the jets reconstruction+labeling efficiencies as obtained for the $WH, H \to b\bar{b}, c\bar{c}, u\bar{u}$ and $gg \to H \to \tau\tau$ events. Jets labeling is optional, can be switched off for b- and c-jets and/or separately for tau-jets (default: ON).

Table 1: *Efficiency for jet reconstruction+labeling for different types of initial partons with* $p_T^{parton} > 15$ *GeV (required* $p_T^{jet} > 15$ *GeV). The rapidity coverage is limited to* $|\eta| < 2.5$*. The* $\Delta R_{cone} = 0.4$ *is used for cluster reconstruction and* $\Delta R_{cone} = 0.2$ *is used for matching criteria. The* $WH, H \to b\bar{b}, c\bar{c}, u\bar{u}$ *and* $gg \to H \to \tau\tau$ *processes were generated with* $m_H = 100$ *GeV. In case of tau-jets only hadronic tau decays were generated.*

| Parton type | Reconstruction + Labeling |
|---|---|
| u-quark | 95% |
| b-quark | 81% |
| c-quark | 87% |
| tau-jet | 80% |

### 2.5.2 Calibration

The reconstructed jets four-momenta need to be corrected for the out-cone energy loss (cascade outside the jet-cone) and for the loss of the particles escaping detection (those below threshold at $p_T = 0.5$ GeV, neutrinos, invisible particles, muons outside acceptance range or below the observability threshold). Such correction, called calibration, can be performed on the statistical basis only. The single default calibration function (the same for any type of jets), as a function of transverse momenta of reconstructed jet, $p_T^{jet}$, is provided in the package. The calibration algorithm corrects jets four-momenta without altering their direction. Calibration can be switched off (default: ON).

The quality of the calibration algorithm can be verified by monitoring the ratio of $p_T^{b-quark}/p_T^{b-jet}$, taking into account a hard-process quark which originates a given jet. One can note, (see Fig. 5, bottom plots) that the implemented calibration function has a tendency to undercalibrate b-jets while calibrates reasonably well u-jets. We can observe also rather large tail in the $p_T^{jet}/p_T^{b-quark}$ distribution, caused by the semileptonic b-quarks decays and larger spread of the cascading particles than in the case of light jets.

### 2.5.3 Reconstruction of the resonance

As an effect of the hadronisation and cascading decays, the expected resolution for the resonance reconstruction in the hadronic channels will be much worse than in the leptonic ones. The precision for reconstructing the peak position of the invariant mass of the di-jet system will relay on the precision of the calibration procedure. Fig. 6 shows two examples of the reconstruction of invariant mass of the di-jet system in the $WH$ production with $H \to u\bar{u}$ and $H \to b\bar{b}$ decays. The long tail at the lower side of the invariant mass distribution is an effect of hadronisation and cascading decays which was not sufficiently corrected by the imposed calibration. One may also notice the effect of the ISR/FSR originated jets contributing to the distribution in the low mass range for the $H \to u\bar{u}$ case where jets identification is more ambiguous.

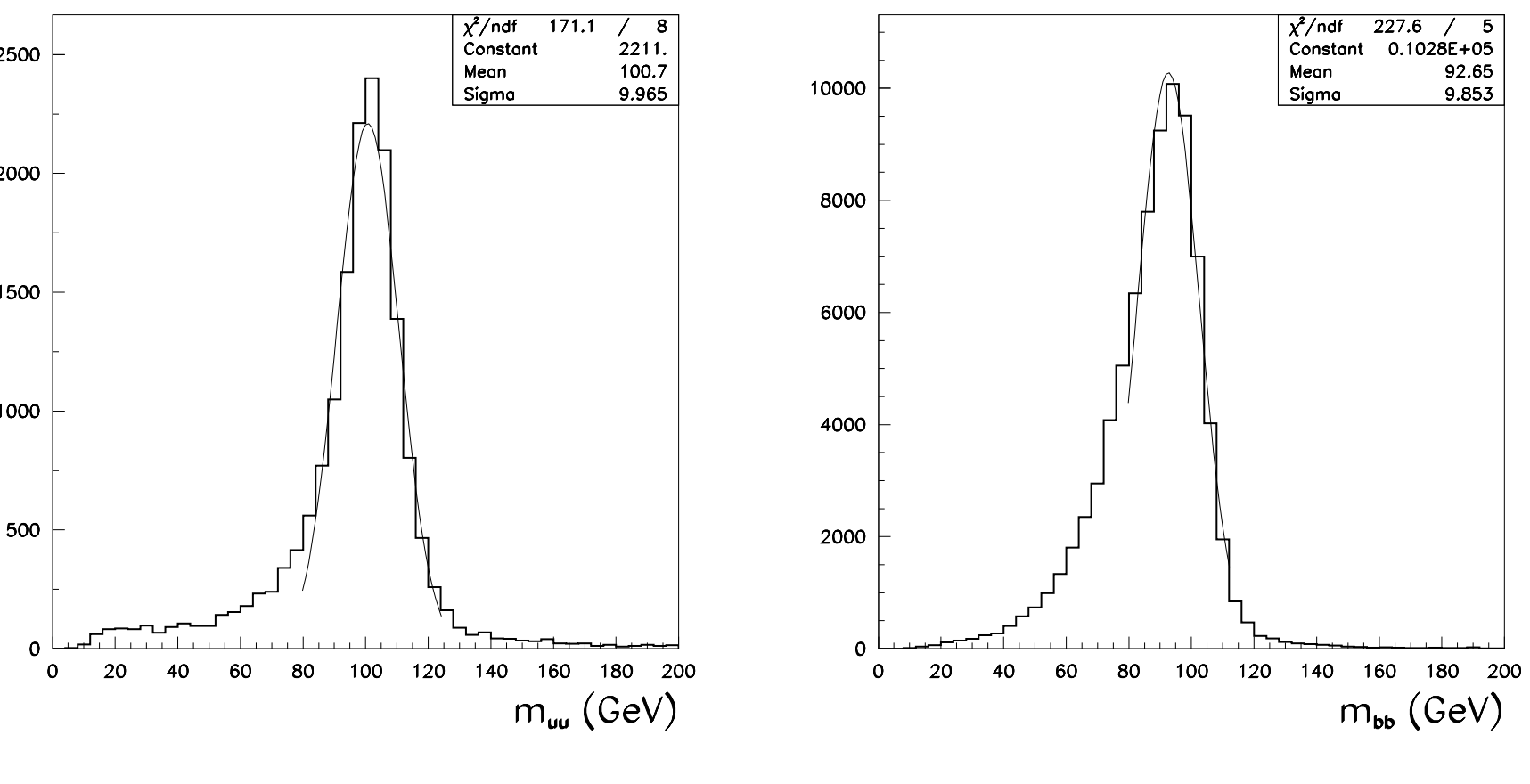

Figure 6: *Left: The reconstructed mass, $m_{jj}$, for the $WH$ with $H \to u\bar{u}$ (left) and $H \to b\bar{b}$ (right) events with $m_H = 100$ GeV.*

## 2.6 Missing transverse energy

The missing transverse energy is calculated by summing the transverse momenta of identified isolated photons, electrons and muons, of jets and clusters not accepted as jets and of non-isolated muons not added to any jet. Finally, the transverse energies deposited in cells not used for clusters reconstruction are also included in the total sum. Transverse energies deposited in unused cells are smeared with the same energy resolution function as for jets, and cells with deposited transverse energy below a given threshold (default: 0 GeV) are excluded from the sum. From the calculation of the total sum $E_T^{obs}$ the missing transverse energy is obtained, $E_T^{miss} = E_T^{obs}$ as well as the missing transverse momentum components $p_x^{miss} = -p_x^{obs}$, $p_y^{miss} = -p_y^{obs}$. The total calorimeter transverse energy, $\sum E_T^{calo}$, is calculated as the sum of all the above transverse energies except that of muons. Please note, that missing transverse energy is calculated from the energy balance before jets calibration is performed, as the possible out-cone energy loss is already taken into account by summing up energy deposition of unused cells/clusters.

Fig. 7 shows the resolution of the transverse missing energy obtained for the di-jet events generated with the transverse momenta of the hard process above 17 GeV.

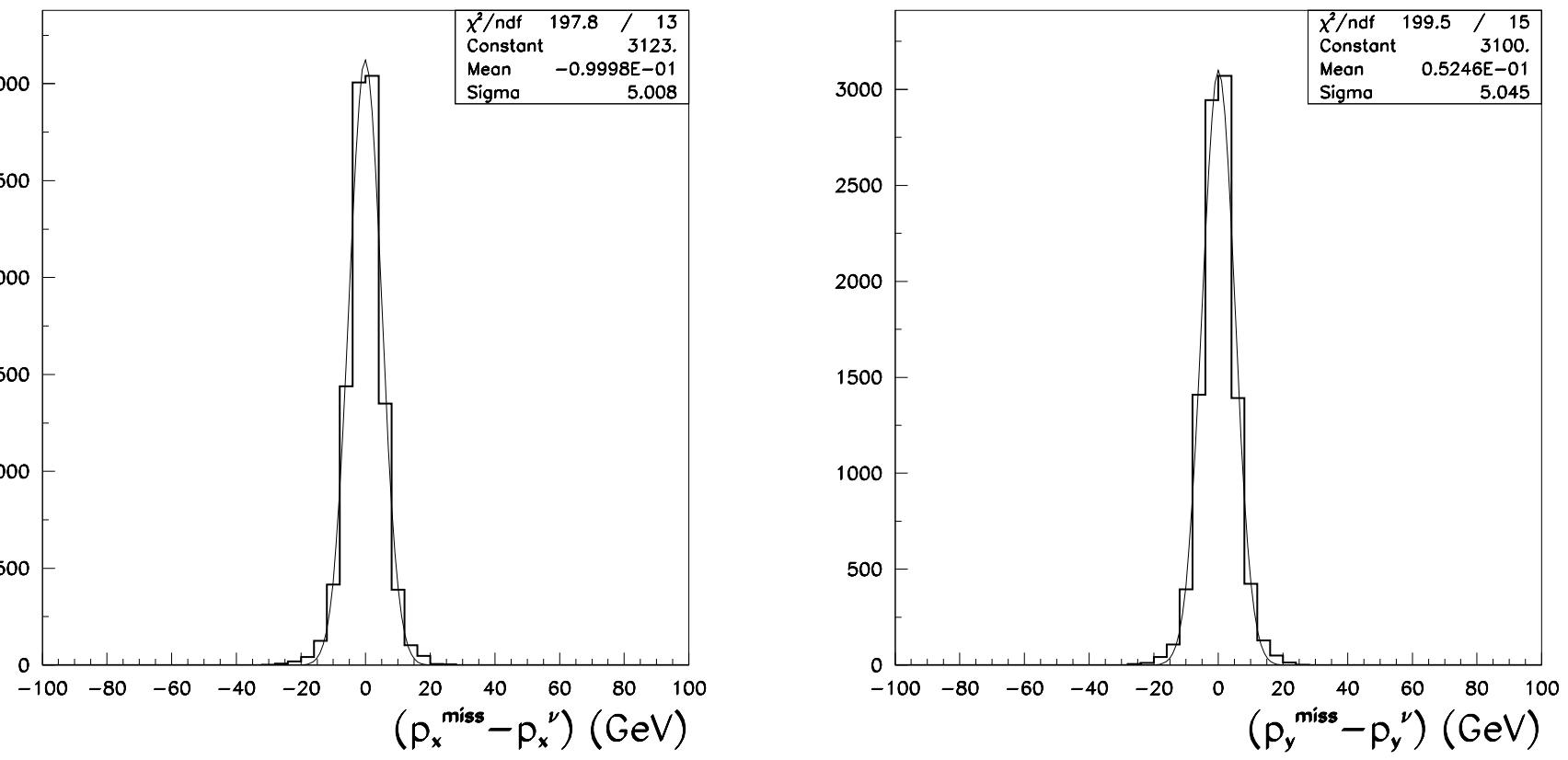

Figure 7: *The $p_x^{miss}$ and $p_y^{miss}$ resolutions for the di-jet events.*

## 2.7 Additional efficiencies

The algorithms of the `AcerDET` package are not correcting for inefficiencies of photon, electron and muon reconstruction and identification. To be more realistic, one should apply a weighting factor of 70%-90% for each isolated lepton used in the analysis and of 80% for each isolated photon.

The package is also not correcting for tagging-efficiencies, namely the labeling procedure is not equivalent to the b- and tau-jet identification in the experiment. One can assume b-tagging efficiency of 60% per b-labeled jet with mistagging probability of 10% for c-labeled jet and 1% for the light jet. For the tau-jets using efficiency of 50% (per tau-labelled jet) and 5 - 10 % mistagging probability for other jets could be a reasonable assumptions. One should be well aware that what proposed above represents quite crude estimates.

The package is also not correcting for trigger efficiencies which has to be applied in addition if a given reconstructed object is foreseen to trigger an event.

### 2.8 OUTPUT format

By the end of the simulation and reconstruction algorithm, reconstructed entities: photons, electrons, muons, jets, transverse missing energy are rewritten to the set of `COMMON BLOCKS`. Provided is algorithm to store this information in the `PAW HBOOK` [6] data-base, so called `PAW NTUPLE`. However, the user might decide to use his preferred data-base technology for storing output from the simulation and reconstruction algorithms.

## 3 Outlook

We presented package which can be useful for several phenomenological studies on the high $p_T$ physics at LHC. One of the favoured application might be the comparison studies on the matrix element and parton shower Monte Carlo predictions for different background processes at LHC, like eg. presented in [5]. It is not the aim of the package to represent in details performance of neither ATLAS nor CMS detectors, nevertheless some global features of these would be reproduced well. In particular we believe that the analyses performed with the package for physics at LHC will be more realistic than parton-level studies alone. The package allows also for rather flexible adjusting of several key parameters which characterise features of any detector for LHC or LC experiments.

## Acknowledgments

This work was inspired by the several years of my involvement in the activity of the Physics Working Groups of the ATLAS Collaboration. I am grateful to all colleagues for a very creative atmosphere. In particular for several suggestions and inspiring discussions to Daniel Froidevaux, who some years ago initiated and guided my work on the first version of the fast simulation package.

## A AcerDET versus ATLAS detector fast and full simulation

Below we compare few benchmarking numbers obtained with the `AcerDET` code, the fast simulation of ATLAS detector `ATLFAST` [2] and the results from the full simulation/reconstruction expected with ATLAS detector [1] at the luminosity of $10^{33} cm^{-2} s^{-1}$. Numbers are given after event selection as foreseen for the analysis in the specific channel [1].

Table 2: *Benchmarking numbers for fast simulation* `AcerDET`*,* `ATLFAST` *package and results from full simulation of ATLAS detector at the luminosity of* $10^{33} cm^{-2} s^{-1}$

| Parton type | AcerDET $\sigma_m$ | ATLFAST $\sigma_m$ | ATLAS full sim. $\sigma_m$ |
|---|---|---|---|
| $H \to \gamma\gamma$ ($m_H = 100$ GeV) | 0.85 GeV | 1.0 GeV | 1.1 GeV |
| $H \to ZZ \to 4e$ ($m_H = 130$ GeV) | 1.26 GeV | 1.45 GeV | 1.5 GeV |
| $H \to ZZ \to 4\mu$ ($m_H = 130$ GeV) | 1.57 GeV | 1.34 GeV | 1.4 GeV |
| $H \to ZZ \to 2e2\mu$ ($m_H = 130$ GeV) | 1.40 GeV | 1.40 GeV | 1.5 GeV |
| $WH, H \to b\bar{b}$ ($m_H = 100$ GeV) | 10 GeV | 12.5 GeV | 14.7 GeV |
| $E_T^{miss}$ resolution (di-jet events $p_T^{hard} > 17$ GeV) | 5.0 GeV | 5.7 GeV | 6.3 GeV |

## B General infomations

The simulation and reconstruction algorithm is executed by a call to the routine `ACERDET`. The events which is going to be processed should be stored in the `COMMON /ACMCEVENT/`:

```
INTEGER N, K
REAL P, V
COMMON/ACMCEVENT/N,K(10000,5),P(10000,5),V(10000,5)
```

The convention for the particles status, mother-daughter relations, particles codes, etc. should be the same as in the `COMMON /PYJETS/` of the `PYTHIA 6.2` event generator [7].

The input/output logical identifiers should be defined in `COMMON /ACDNOUT/`:

```
INTEGER NINP,NOUT
COMMON /ACDNOUT/ NINP,NOUT
```

The package reads single input file `acerdet.dat` which contains parameters for events simulation and reconstruction and write out single control output file `acerdet.out`. If initialised by the user in the main program, control histograms and ntuple with reconstructed events will be stored in the `acerdet.ntup` file.

### B.1 Subroutine ACERDET

This is the main subroutine, called to execute simulation and reconstruction algorithms. The following modes of executing this subroutine are implemented:

- `MODE=-1` – initalisation, which should be called before the first event is processed.
- `MODE= 0` – simulation and reconstruction, should be called event by event.
- `MODE= 1` – finalisation, should be called after last event is processed.

This subroutines invokes other subroutines: `ACDINF`, `ACDINI`, `ACDCEL`, `ACDCLU`, `ACDELE`, `ACDMUO`, `ACDPHO`, `ACDJET`, `ACDMIS`, `ACDBJE`, `ACDCJE`, `ACDTAU` and `ACDCAL`.

### B.2 Interfaces to Event Generators

The event, which is going to be processed, should be stored in the `COMMON /ACMCEVENT/`. In the source code of the program provided are subroutines `ACDPYTHIA6` and `ACDHERWIG6` which respectively fills the `COMMON /ACMCEVENT/` with the event content from `COMMON /PYJETS/` (for `PYTHIA` generator) and `COMMON /HEPEVT/` (for `HERWIG` generator [8]). In the second case also the convention for status of the particles is modified and made consistent with that used by the `ACERDET`.

### B.3 External calling sequence

The following calling sequence should be provided in the main program

```
Initialise Generator
CALL ACERDET(-1)
CALL ACERDETNTUP(-1)
DO II= 1,NEVENT
   Generate Event (PYTHIA) or  Generate Event (HERWIG)
   CALL ACDPYTHIA6 or  CALL ACDHERWIG6
   CALL ACERDET(0)
   CALL ACERDETNTUP(0)
ENDDO
Finalise Generator
CALL ACERDET(1)
CALL ACERDETNTUP(1)
```

Invoking subroutine `ACERDETNTUP` is optional. It stores data-base with processed events in the `HBOOK NTUPLE` [6] from `cernlib` library. It creates also output file for storing control histograms and ntuple in `acerdet.ntup`. User however may decide on a different output format.

If user do not wish, the dependencies on `cernlib` could be easily removed. It requires linking into executable the respective library and providing very simple conversion package where the existing calls to `cernlib` functions for random number generation and histograming are used to fill information into the user's preferred ones.

### B.4 Execution sequence

After the initialisation phase, the event by event execution sequence is a following one:

- Map of the cells energy deposition is created: `SUBROUTINE ACDCEL`.
- Calorimetric clusters are reconstructed: `SUBROUTINE ACDCLU`.
- Isolated and non-isolated muons are reconstructed: `SUBROUTINE ACDMUO`.
- Isolated electrons are reconstructed: `SUBROUTINE ACDELE`.
- Isolated photons are reconstructed: `SUBROUTINE ACDPHO`.
- The remaining calorimetric clusters are identified as jets: `SUBROUTINE ACDJET`.
- Missing transverse is calculated for the reconstructed event: `SUBROUTINE ACDMIS`.
- Optionally, if specified in the acerdet.dat file, algorithms for jets labeling and calibration of jets are executed (default=ON): `SUBROUTINE ACDBJE, SUBROUTINE ACDCJE, SUBROUTINE ACDTAU, SUBROUTINE ACDCAL`
- Reconstructed event is stored in the final common blocks: `SUBROUTINE ACDOUT`.

The respective sequence of calls is executed in the `SUBROUTINE ACERDET`.

## B.5 Structure of the distribution version

The distribution version consists of the source code of the `AcerDET` and example of the main program for execution with `PYTHIA 6.2` generator. The `acerdet.dat` file resides in subdirectory `acerdet_dat`. Source code resides in `acerdet_src` with includes files stored in `acerdet_inc` The library will be created in `acerdet_lib`. provided is also example of the main program `demo.f` to execute `AcerDET` algorithms with `PYTHIA 6.2` generator.

```
-rw-rw-r--    1 erichter erichter     1613 Jul 30 08:46 AAREADME
drwxr-xr-x    2 erichter erichter     4096 Jul 29 20:17 acerdet_dat
drwxr-xr-x    2 erichter erichter     4096 Jul 29 20:17 acerdet_inc
drwxrwxr-x    2 erichter erichter     4096 Jul 29 20:21 acerdet_lib
drwxr-xr-x    2 erichter erichter     4096 Jul 29 20:21 acerdet_src
-rw-r--r--    1 erichter erichter    10682 Jul  5 16:48 demo.f
-rw-r--r--    1 erichter erichter     1337 Jul 29 20:23 makefile
drwxr-xr-x    2 erichter erichter     4096 Jul 29 20:24 prod
```

To execute the package following actions should be taken:

- In subdirectory `acerdet_src` type: *make install.* It will compile the code and create the `libacerdet.a` library in `acerdet_lib` subdirectory
- Edit `makefile` in main directory to define `PYPATH` and `CERNLIB` environmental variables
- In main directory type: *make*. It will create `demo.exe` executable.
- Optional: Edit `run.dat` file in subdirectory `prod` to redefine generated process, number of events, random seed, etc.
- In main directory type: *make run.* It will create link to the `acerdet.dat` file in `prod` subdirectory and will execute the executable. Output will be stored in subdirectory `prod`.

## C List of input parameters: acerdet.dat file

```
C------ flags and switches ---------------
     10000       ---  LPAR(1)   ....id for histograms
         1       ---  LPAR(2)   ....smearing on=1, off=0
         1       ---  LPAR(3)   ....B-field  on=1, off=0
        66       ---  LPAR(4)   ....code for invisible particle
         1       ---  LPAR(5)   ....b- and c-labeling  on=1, off=0
         1       ---  LPAR(6)   ....tau-labeling  on=1, off=0
         1       ---  LPAR(7)   ....calibration   on=1, off=0
C------ parameters for ACDCEL ---------------
     5.000       ---  YPAR(1)  ....rapidity coverage
     0.500       ---  YPAR(2)  ....min p_T for B-field
     0.000       ---  YPAR(3)  ....min E_T for cell
     3.200       ---  YPAR(4)  ....eta transition in cells granularity
     0.100       ---  YPAR(5)  ....granularity in eta (within YPAR(17)), 2x outside
     0.100       ---  YPAR(6)  ....granularity in phi (within YPAR(17)), 2x outside
C------ parameters for ACDCLU ---------------
     5.000       ---  YPAR(10)  ....minimum E_T for cluster
     0.400       ---  YPAR(11)  ....cone R for clustering
     5.000       ---  YPAR(12)  ....rapidity coverage
     1.500       ---  YPAR(13)  ....min E_T for cluster initiator
C------ parameters for ACDMUO ---------------
     6.000       ---  YPAR(20)  ....minimum muon-momenta to be detected
     2.500       ---  YPAR(21)  ....maximum muon eta to be detected
     0.400       ---  YPAR(27)  ....min R_lj for muon-isolation
     0.200       ---  YPAR(28)  ....R_cone for energy deposition
    10.000       ---  YPAR(29)  ....max energy deposition for isol
C------ parameters for ACDPHO ---------------
     5.000       ---  YPAR(30)  ....minimum photon-momenta to be isol
     2.500       ---  YPAR(31)  ....maximum photon eta to be isol
     0.150       ---  YPAR(35)  ....min R_lj for photon-jet
     0.400       ---  YPAR(37)  ....min R_lj for photon-isolation
     0.200       ---  YPAR(38)  ....R_cone for energy deposition
    10.000       ---  YPAR(39)  ....max energy deposition for isol
C------ parameters for ACDELE ---------------
     5.000       ---  YPAR(40)  ....minimum electron-momenta to be isol
     2.500       ---  YPAR(41)  ....maximum electron eta to be isol
     0.150       ---  YPAR(45)  ....min R_lj for electron-jet
     0.400       ---  YPAR(47)  ....min R_lj for electron-isolation
     0.200       ---  YPAR(48)  ....R_cone for energy deposition
    10.000       ---  YPAR(49)  ....max energy deposition for isol
C------ parameters for ACDJET ---------------
    10.000       ---  YPAR(51)  ....jets energy_min  threshold
     5.000       ---  YPAR(52)  ....rapidity coverage for jets
C------ parameters for ACDBJE ---------------
     5.000       ---  YPAR(61)  ....minimum b-quark pT (after FSR) momenta for b-jet label
     2.500       ---  YPAR(62)  ....maximum b-quark eta for  b-jet label
     0.200       ---  YPAR(63)  ....max R_bj for b-jet label
C------ parameters for ACDCJE ---------------
     5.000       ---  YPAR(71)  ....minimum c-quark pT (after FSR) momenta for c-jet label
     2.500       ---  YPAR(72)  ....maximum c-quark eta for  c-jet label
     0.200       ---  YPAR(73)  ....max R_cj for c-jet label
C------ parameters for ACDTAU ---------------
    10.000       ---  YPAR(90)  ....minimum tau-had pT for  tau-jet label
     2.500       ---  YPAR(91)  ....maximum tau-eta for  tau-jet label
     0.300       ---  YPAR(92)  ....max R_tauj for tau-jet
     0.900       ---  YPAR(93)  ....max R_tauj for tau-jet
C------ parameters for ACDMIS ---------------
     0.000       ---  YPAR(80)  ....min E_T for energy in cell to count unused cell
C------   end of data files   ---------------
```

# D Reconstructed entities

Isolated photons are stored in `COMMON /ACDPHOT/`, given are 4-momenta in the laboratory frame `PXPHOT, PYPHOT, PZPHOT, EEPHOT`, and code for the photon `KFPHOT=22`. Photons are ordered with the decreasing transverse momenta.

```
      INTEGER MAXPHOT
      PARAMETER(MAXPHOT=12)
      INTEGER NPHOT,KFPHOT
      REAL PXPHOT,PYPHOT,PZPHOT,EEPHOT
      COMMON /ACDPHOT/ NPHOT,KFPHOT(MAXPHOT),
     +                 PXPHOT(MAXPHOT),PYPHOT(MAXPHOT),PZPHOT(MAXPHOT),
     +                 EEPHOT(MAXPHOT)
```

Isolated leptons are stored in `COMMON /ACDLEPT/`, given are 4-momenta in the laboratory frame `PXLEPT, PYLEPT, PZLEPT, EELEPT`, and code for the lepton `KFLEPT=`$\pm$ `11`, $\pm$ `13`. Leptons are ordered with the decreasing transverse momenta.

```
      INTEGER MAXLEPT
      PARAMETER(MAXLEPT=12)
      INTEGER NLEPT,KFLEPT
      REAL PXLEPT,PYLEPT,PZLEPT,EELEPT
      COMMON /ACDLEPTT/ NLEPT,KFLEPT(MAXLEPT),
     +                 PXLEPT(MAXLEPT),PYLEPT(MAXLEPT),PZLEPT(MAXLEPT),
     +                 EELEPT(MAXLEPT)
```

Reconstructed jets stored in `COMMON /ACDJETS/`, given are 4-momenta in the laboratory frame `PXJETS, PYJETS, PZJETS, EEJETS`, and code for the jet `KFJETS= 5, 4, 15, 98`; respectively for b-jets, c-jets, tau-jets and light-jets. Jets are ordered with the decreasing transverse moment.

```
      INTEGER MAXJET
      PARAMETER(MAXJET=20)
      INTEGER NJETS,KFJETS
      REAL PXJETS,PYJETS,PZJETS,EEJETS
      COMMON /ACDJETS/  NJETS,KFJETS(MAXJET),
     +                  PXJETS(MAXJET),PYJETS(MAXJET),PZJETS(MAXJET),
     +                  EEJETS(MAXJET)
```

Information on the reconstructed missing transverse energy is stored in `COMMON /ACDMISS/`. Given are both transverse components `PXMISS,PYMISS`, sum of transverse momenta components of neutrinos and invisible particles (as defined by the user) present in the event `PXNUES,PYNUES` and finally sum of the total transverse energy components deposited in the calorimeter `PXCALO,PYCALO`.

```
      REAL PXMISS,PYMISS,PXNUES,PYNUES,PXCALO,PYCALO
      COMMON /ACDMISS/PXMISS,PYMISS,PXNUES,PYNUES,PXCALO,PYCALO
```

For the convenience stored is also information on the history of all particles participating in the *hard-scattering* in which event was created. Included are those with the status code = 21 (convention of `PYTHIA` generator). Given is: status of particle `KSPART`, its flavour code `KFPART` and 4-momenta in the laboratory frame `PXPART, PYPART, PZPART, EEPART`.

```
      INTEGER MAXPART
      PARAMETER(MAXPART=40)
      INTEGER NPART,KSPART,KFPART
      REAL PXPART,PYPART,PZPART,EEPART
      COMMON /ACDPART/ NPART,KSPART(MAXPART),KFPART(MAXPART),
     +                 PXPART(MAXPART),PYPART(MAXPART),
     +                 PZPART(MAXPART),EEPART(MAXPART)
```

Some additional information, if required, can be also extracted from the internal variables/common blocks of simulation and reconstruction algorithms: eg. on used/unused cells, non-isolated muons, unused clusters, etc.

# E Parametrisation for energy/momenta resolution

## E.1 Function RESPHO

The parametrisation for photon energy resolution assumes only energy dependence and the Gaussian smearing with $10\%/\sqrt{E_\gamma}$ resolution.

$$E_\gamma^{smeared} = E_\gamma^{true} \cdot (1 + r_n \cdot \frac{0.10}{\sqrt{E_\gamma^{true}}}). \tag{1}$$

Where $r_n$ is the random number generated according to the Gaussian distribution (function `RANNOR` from `cernlib` library is used). All 4-momenta components of the photon are smeared with the same resolution so the direction of the photon is not altered.

## E.2 Function RESELE

The parametrisation for electron energy resolution assumes only energy dependence and the Gaussian smearing with $12\%/\sqrt{E_e}$ resolution.

$$E_e^{smeared} = E_e^{true} \cdot (1 + r_n \cdot \frac{0.12}{\sqrt{E_e^{true}}}). \tag{2}$$

Where $r_n$ is the random number generated according to the Gaussian distribution (function `RANNOR` from `cernlib` library is used). All 4-momenta components of the electron are smeared with the same resolution so the direction of the electron is not altered.

## E.3 Function RESHAD

The parametrisation for clusters momenta resolution assumes only energy dependence and Gaussian smearing with $50\%/\sqrt{E_{clu}}$ or $100\%/\sqrt{E_{clu}}$ resolution. The transition region, $|\eta^{clu}| = CALOTH$ is the same as the transition of the cells granularity.
For $|\eta^{clu}| < CALOTH$:

$$E_{clu}^{smeared} = E_{clu}^{true} \cdot (1 + r_n \cdot \frac{0.50}{\sqrt{E_{clu}^{true}}}). \tag{3}$$

For $|\eta^{clu}| > CALOTH$:

$$E_{clu}^{smeared} = E_{clu}^{true} \cdot (1 + r_n \cdot \frac{1.00}{\sqrt{E_{clu}^{true}}}). \tag{4}$$

Where $r_n$ is the random number generated according to the Gaussian distribution (function `RANNOR` from `cernlib` library is used). All 4-momenta components of the muon are smeared with the same resolution so the direction of the muon is not altered.

### E.4 Function RESMUO

The parametrisation for muon momenta resolution assumes only transverse momenta dependence and Gaussian smearing with $0.0005 \cdot pT_\mu$ resolution.

$$pT_\mu^{smeared} = pT_\mu^{true}/(1 + r_n \cdot 0.0005 \cdot pT_\mu^{true}). \tag{5}$$

Where $r_n$ is the random number generated according to the Gaussian distribution (function `RANNOR` from `cernlib` library is used). All 4-momenta components of the muon are smeared with the same resolution so the direction of the muon is not altered.

### E.5 Function FLDPHI

The effect of the magnetic field is included by simple shifting the $\phi$ position of the charged particles respectively to its transverse momenta, parametrised as following:

$$\delta\phi = 0.5/pT^{part}. \tag{6}$$

The $\delta\phi$ is calculated in radians and $pT^{part}$ is given in $GeV$. The sign of the $\delta\phi$ is the same as the sign of the particle charge. Charged particles with $pT^{part} < 0.5$ GeV are assumed to be looping in the detector and not depositing energy in the calorimeter.

## F Some formulas

We work with the assumptions of massless reconstructed objets. The following relations were used to translate between $(p_T, \eta, \phi)$ coordinates and four-momenta $(p_x, p_y, p_z, E)$.

$$p_x = p_T \cdot cos(\phi) \tag{7}$$

$$p_y = p_T \cdot sin(\phi) \tag{8}$$

$$p_z = p_T \cdot cosh(\eta) \tag{9}$$

$$E = p_T \cdot sinh(\eta) \tag{10}$$

$$p_T = \sqrt{p_x^2 + p_y^2} \tag{11}$$

$$\eta = sign(\ln \frac{\sqrt{p_T^2 + p_z^2} + |p_z|}{p_T}, p_z) \tag{12}$$

$$\phi = asinh(p_y/\sqrt{p_x^2 + p_y^2}) \tag{13}$$

# G Output content

## G.1 Output ntuple content: `acerdet.ntup` file

The content of the ntuple is exactly as described in the previous section. As an auxiliary info, within the stored information, could be provided code on the generated process, `IDPROC`.

```
******************************************************************
* Ntuple ID = 3333   Entries = 100000    ACERDET
******************************************************************
* Var numb * Type * Packing *    Range     *  Block   *  Name    *
******************************************************************
*      1   * I*4  *         *              * ACDINFO  * IDPROC
*      1   * R*4  *         *              * ACDMISS  * PXMISS
*      2   * R*4  *         *              * ACDMISS  * PYMISS
*      3   * R*4  *         *              * ACDMISS  * PXNUES
*      4   * R*4  *         *              * ACDMISS  * PYNUES
*      5   * R*4  *         *              * ACDMISS  * PXCALO
*      6   * R*4  *         *              * ACDMISS  * PYCALO
*      1   * I*4  *         * [0,12]       * ACDLEPT  * NLEPT
*      2   * I*4  *         *              * ACDLEPT  * KFLEPT(NLEPT)
*      3   * R*4  *         *              * ACDLEPT  * PXLEPT(NLEPT)
*      4   * R*4  *         *              * ACDLEPT  * PYLEPT(NLEPT)
*      5   * R*4  *         *              * ACDLEPT  * PZLEPT(NLEPT)
*      6   * R*4  *         *              * ACDLEPT  * EELEPT(NLEPT)
*      1   * I*4  *         * [0,12]       * ACDPHOT  * NPHOT
*      2   * I*4  *         *              * ACDPHOT  * KFPHOT(NPHOT)
*      3   * R*4  *         *              * ACDPHOT  * PXPHOT(NPHOT)
*      4   * R*4  *         *              * ACDPHOT  * PYPHOT(NPHOT)
*      5   * R*4  *         *              * ACDPHOT  * PZPHOT(NPHOT)
*      6   * R*4  *         *              * ACDPHOT  * EEPHOT(NPHOT)
*      1   * I*4  *         * [0,20]       * ACDJETS  * NJETS
*      2   * I*4  *         *              * ACDJETS  * KFJETS(NJETS)
*      3   * R*4  *         *              * ACDJETS  * PXJETS(NJETS)
*      4   * R*4  *         *              * ACDJETS  * PYJETS(NJETS)
*      5   * R*4  *         *              * ACDJETS  * PZJETS(NJETS)
*      6   * R*4  *         *              * ACDJETS  * EEJETS(NJETS)
*      1   * I*4  *         * [0,40]       * ACDPART  * NPART
*      2   * I*4  *         *              * ACDPART  * KSPART(NPART)
*      3   * I*4  *         *              * ACDPART  * KFPART(NPART)
*      4   * R*4  *         *              * ACDPART  * PXPART(NPART)
*      5   * R*4  *         *              * ACDPART  * PYPART(NPART)
*      6   * R*4  *         *              * ACDPART  * PZPART(NPART)
*      7   * R*4  *         *              * ACDPART  * EEPART(NPART)
******************************************************************
*  Block   *  Entries  * Unpacked * Packed *   Packing Factor    *
******************************************************************
* ACDINFO  *  100000   * 4        * 4      *       1.000         *
* ACDMISS  *  100000   * 24       * 24     *       1.000         *
* ACDLEPT  *  100000   * 244      * Var.   *    Variable         *
* ACDPHOT  *  100000   * 244      * Var.   *    Variable         *
* ACDJETS  *  100000   * 404      * Var.   *    Variable         *
* ACDPART  *  100000   * 964      * Var.   *    Variable         *
* Total    *    ---    * 1884     * Var.   *    Variable         *
******************************************************************
* Blocks = 6            Variables = 32      Max. Columns = 471   *
******************************************************************
```

## G.2 Control printout: acerdet.out file

```
***************************************************************************
*                                                                         *
*                    *************************                            *
*                     AcerDET, version: 1.0                               *
*                     Released at: 30.07.2002                             *
*                    *************************                            *
*                                                                         *
*          Simplied event simulation and reconstruction package           *
*                                                                         *
*                       by E. Richter-Was                                 *
*                  Institute of Computer Science                          *
*               Jagellonian University, Cracow, Poland                    *
*                                                                         *
*                                                                         *
***************************************************************************
***************************************************************************
*               ********************************                          *
*               *     ****************         *                          *
*               *     ***  ACDCEL   ***        *                          *
*               *     ****************         *                          *
*               ********************************                          *
*                 clusters definition ....                                *
*          5.00000               eta coverage              ETACEL         *
*          0.00000               E_T_min cell thresh       ETTHR          *
*          3.20000               eta gran. transition      CALOTH         *
*          0.10000               gran in eta(central)      DBETA          *
*          0.10000               gran in phi(central)      DBPHI          *
*                 B field apply ....                                      *
*                1               B-field on/off            KEYFLD         *
*          0.50000               p_T min non looping       PTMIN          *
*                 invisible particles ....                                *
*               66               KF code for invis         KFINVS         *
***************************************************************************
***************************************************************************
*               ********************************                          *
*               *     ****************         *                          *
*               *     ***  ACDCLU   ***        *                          *
*               *     ****************         *                          *
*               ********************************                          *
*                 clusters definition ....                                *
*          5.00000               E_T_min cluster          ETCLU           *
*          1.50000               E_T_min cell initia      ETINI           *
*          0.40000               R cone                   RCONE           *
*          5.00000               eta coverage             ETACLU          *
*          3.20000               eta gran. transition     CALOTH          *
***************************************************************************
***************************************************************************
*               ********************************                          *
*               *     ****************         *                          *
*               *     ***   ACDMUO   ***       *                          *
*               *     ****************         *                          *
*               ********************************                          *
*                 muon isolation ....                                     *
*          6.00000               min. muon p_T            PTMUMIN         *
*          2.50000               max. muon eta            ETAMAX          *
*          0.40000               min R_lj for isolat.     RISOLJ          *
*          0.20000               R for energy deposit     RDEP            *
*         10.00000               max E_dep for isolat     EDMAX           *
*          0.00000               min. muon p_T unsmea     PTMUMIT         *
*                1               smearing on/off          KEYSME          *
***************************************************************************
```

```
****************************************************************************
*               ********************************                           *
*               *     *****************        *                           *
*               *     ***   ACDELE  ***        *                           *
*               *     *****************        *                           *
*               ********************************                           *
*                 electron isolation ....                                  *
*          5.00000                min. lepton p_T          PTLMIN          *
*          2.50000                max. lepton eta          ETAMAX          *
*          0.15000                max R_ej for ele-clu     RJE             *
*          0.40000                min R_lj for isolat.     RISOLJ          *
*          0.20000                R for energy deposit     RDEP            *
*         10.00000                max E_dep for isolat     EDMAX           *
*                1                smearing on/off          KEYSME          *
****************************************************************************
****************************************************************************
*               ********************************                           *
*               *     *****************        *                           *
*               *     ***   ACDPHO   ***       *                           *
*               *     *****************        *                           *
*               ********************************                           *
*                 photon  isolation ....                                   *
*          5.00000                min. photon p_T          PTLMIN          *
*          2.50000                max. photon eta          ETAMAX          *
*          0.15000                max R_gam-clust          RJE             *
*          0.40000                min R_isol               RISOLJ          *
*          0.20000                R for energy deposit     RDEP            *
*         10.00000                max E_dep for isolat     EDMAX           *
*                1                smearing on/off          KEYSME          *
****************************************************************************
****************************************************************************
*               ********************************                           *
*               *     *****************        *                           *
*               *     ***  ACDJET   ***        *                           *
*               *     *****************        *                           *
*               ********************************                           *
*                 clusters definition ....                                 *
*          0.40000                R cone                    RCONE          *
*                 jets definition ....                                     *
*         10.00000                E_T_jets [GeV]            ETJET          *
*          5.00000                eta coverage jets         ETAJET         *
*                1                smearing on/off           KEYSME         *
****************************************************************************
****************************************************************************
*               ********************************                           *
*               *     *****************        *                           *
*               *     ***   ACDMIS   ***       *                           *
*               *     *****************        *                           *
*               ********************************                           *
*                    muon coverage     ....                                *
*          6.00000                min. muon p_T             PTMUMIN        *
*          2.50000                max. muon eta             ETAMAX         *
*                    unused cells     ....                                 *
*                1                smearing on/off           KEYSME         *
*          0.00000                cells threshold           ETCELL         *
*                    invisible particles     ....                          *
*               66                KF code for invis         KFINVS         *
****************************************************************************
```

```
***************************************************************************
*              ********************************                           *
*              *     ****************        *                            *
*              *     ***   ACDBJE   ***      *                            *
*              *     ****************        *                            *
*              ********************************                           *
*                  jets labeling ....                                     *
*               1       labeling on/off                   KEYBCL          *
*                  jets definition ....                                   *
*        10.00000               E_T_jets [GeV]            ETJET           *
*                  b-jets............ ....                                *
*         5.00000               min b-quark p_T           PTBMIN          *
*         2.50000               max b-quark eta           ETBMAX          *
*         0.20000               max R_bj for b-jet        RJB             *
***************************************************************************
***************************************************************************
*              ********************************                           *
*              *     ****************        *                            *
*              *     ***   ACDCJE   ***      *                            *
*              *     ****************        *                            *
*              ********************************                           *
*                jets labeling ....                                       *
*               1       labeling on/off                   KEYBCL          *
*                jets definition ....                                     *
*        10.00000               E_T_jets [GeV]            ETJET           *
*                c-jets............                                       *
*         5.00000               min c-quark p_T           PTCMIN          *
*         2.50000               max c-quark eta           ETCMAX          *
*         0.20000               max R_cj for c-jet        RJC             *
***************************************************************************
***************************************************************************
*              ********************************                           *
*              *     ****************        *                            *
*              *     ***   ACDTAU   ***      *                            *
*              *     ****************        *                            *
*              ********************************                           *
*                jets labeling ....                                       *
*               1       labeling on/off                   KEYBCL          *
*                jets definition ....                                     *
*        10.00000               E_T_jets [GeV]            ETJET           *
*                tau-jets............                                     *
*        10.00000               min tau-had p_T           PTTAU           *
*         2.50000               max tau-had eta           ETATAU          *
*         0.30000               max R_tauj for tau-j      RJTAU           *
*         0.90000               tau-had frac. of jet      PTFRAC          *
***************************************************************************

***************************************************************************
*              ********************************                           *
*              *     ****************        *                            *
*              *     ***   ACDCAL   ***      *                            *
*              *     ****************        *                            *
*              ********************************                           *
*               jets calibration ....                                     *
*               1               calibration n/off         KEYCAL          *
***************************************************************************
```

## G.3 Control histograms: acerdet.ntup file

Below is the list of control histograms which monitors performance of the simulation/reconstruction. The are store together with ACERDET ntuple in file acerdet.ntup.

```
10001 (1)   ACDCEL: cells multiplicity
10101 (1)   ACDCLU: clusters multiplicity
10111 (1)   ACDCLU: delta phi clu-barycentre
10112 (1)   ACDCLU: delta eta clu-barycentre
10113 (1)   ACDCLU: delta r clu-barycentre
10123 (1)   ACDCLU: delta r clu-parton
10114 (1)   ACDCLU: pTclu/SumpTparticle
10124 (1)   ACDCLU: pTclu/pTparton
10210 (1)   ACDMUO: muon multiplicity NOISOLATED
10211 (1)   ACDMUO: muon multiplicity ISOLATED
10221 (1)   ACDMUO: muon multiplicity HARD
10231 (1)   ACDMUO: muon multiplicity HARD+isol
10311 (1)   ACDELE: electron multiplicity ISOLATED
10321 (1)   ACDELE: electron multiplicity HARD
10331 (1)   ACDELE: electron multiplicity HARD+isol
10411 (1)   ACDPHO: photon multiplicity ISOLATED
10421 (1)   ACDPHO: photon multiplicity HARD
10431 (1)   ACDPHO: photon multiplicity HARD+ISO
10501 (1)   ACDJET: jets multiplicity
10511 (1)   ACDJET: delta phi jet-barycentre
10512 (1)   ACDJET: delta eta jet-barycentre
10513 (1)   ACDJET: delta r jet-barycentre
10523 (1)   ACDJET: delta r jet-parton
10514 (1)   ACDJET: pTjet/SumpTparticle
10524 (1)   ACDJET: pTjet/pTparton
10611 (1)   ACDMIS: reconstructed p_T
10612 (1)   ACDMIS: reconstructed p_T +cells
10613 (1)   ACDMIS: pTmiss
10621 (1)   ACDMIS: p_T nu
10711 (1)   ACDBJE: b-jets multiplicity
10721 (1)   ACDBJE: b-quarks HARD multiplicity
10723 (1)   ACDJET: delta r bjet-bquark
10724 (1)   ACDJET: pTbjet/pTbquark
10811 (1)   ACDCJE: c-jets multiplicity
10821 (1)   ACDCJE: c-quarks HARD multiplicity
10823 (1)   ACDJET: delta r cjet-cquark
10824 (1)   ACDJET: pTcjet/pTcquark
10911 (1)   ACDTAU: tau-jets multiplicity
10921 (1)   ACDTAU: taus  multiplicity
11010 (1)   ACDCAL: calibration entry
11011 (1)   ACDCAL: calibration factor
11020 (1)   ACDCAL: pTjetCALIB/pTparton
11030 (1)   ACDCAL: pTjetNONCALIB/pTparton
```